\documentclass[aip,pof,reprint]{revtex4-1}
\usepackage[T1]{fontenc}
\usepackage[utf8]{inputenc}
\usepackage{verbatim}
\usepackage{amsmath}
\usepackage{graphicx}
\usepackage{hyperref}
\usepackage{psfrag}
\usepackage{stmaryrd}
\usepackage{amssymb}
\usepackage{wasysym}
\usepackage{hyperref}
\usepackage{ulem}
\usepackage[dvips,usenames,dvipsnames]{color}

\begin{document}

\title{Experimental Parametric Subharmonic Instability in Stratified Fluids}

\author{Sylvain Joubaud}
\email{sylvain.joubaud@ens-lyon.fr}
\affiliation{Laboratoire de Physique de l'\'Ecole Normale Sup\'erieure de Lyon, CNRS and Universit\'e de Lyon, 46 All\'ee d'Italie, 69007 Lyon, France}

\author{James Munroe}
\email{jmunroe@mun.ca}
\affiliation{Department of Physics and Physical Oceanography, Memorial University of Newfoundland, St. John's, NL A1B 3X7, Canada}

\author{Philippe Odier}
\email{philippe.odier@ens-lyon.fr}
\affiliation{Laboratoire de Physique de l'\'Ecole Normale Sup\'erieure de Lyon, CNRS and Universit\'e de Lyon, 46 All\'ee d'Italie, 69007 Lyon, France}

\author{Thierry Dauxois}
\email{thierry.dauxois@ens-lyon.fr}
\affiliation{Laboratoire de Physique de l'\'Ecole Normale Sup\'erieure de Lyon, CNRS and Universit\'e de Lyon, 46 All\'ee d'Italie, 69007 Lyon, France}

\date{\today}

\begin{abstract}
Internal gravity waves contribute to fluid mixing and energy transport, not only in oceans but also in the atmosphere and in astrophysical bodies. An efficient way to transfer energy from large scale to smaller scale is the parametric subharmonic instability. We provide here the first experimental measurement of the  growth rate of this instability. We make careful and quantitative comparisons with theoretical predictions  for propagating vertical modes in laboratory experiments.
\end{abstract}


\maketitle

Internal gravity waves (IGW) result from the balance of inertia and buoyancy forces in a density stratified fluid.
Such waves have received a great deal of attention recently because of their relevance and ubiquity in different
physical situations: they are believed to be of primary importance as they affect ocean mixing 
and energy transport~\cite{Kunze}.   Although internal gravity waves do not 
play the dominant role in the evolution of weather and climate, their influence is non-negligible in the dynamics 
of the atmosphere~\cite{BruceBook}. 
From  a fundamental point of view, these waves are also particularly intriguing.
A striking consequence of stratification 
is an anisotropic dispersion relation relating the frequency to the direction of propagation of the wave  
and not to the wavelength. This  property is also encountered for inertial waves (in presence of rotation) or
plasma waves (in presence of a magnetic field). This has unexpected and interesting consequences in the propagation, 
reflection~\cite{PRLSwinney} or transmission properties of these waves~\cite{PRLManiTom}.

Internal waves are known to be inherently unstable due to parametric subharmonic instability (PSI)~\cite{Staquet02}. PSI is a type of resonant triad interaction  where nonlinear terms in the equations of motion allow for efficient transfer of energy from large to small length scales where it can be dissipated. The terminology "parametric sub-harmonic" is used because, for inviscid fluids, PSI transfers energy from a primary wave to two recipient waves of half the frequency. As viscosity effects set in, the frequencies of the recipient waves diverge from half the frequency of the primary wave. In previous laboratory experiments, PSI has been qualitatively observed by driving low-order standing modes with plungers on
the sides of the container~\cite{Thorpe69},  with  an oscillating paddle~\cite{McEwan71,McEwan1972}
or relying on the parametric forcing of the tank~\cite{Benielli98}. For large amplitude forcing, ``irregularities'' or ``traumata'' were observed, which led to mixing and overturning.  In ref~\cite{McEwan1972}, the critical amplitude of the instability has been measured. Quantitative measurements of the growth rate of the instability have never been reported. 

We present here experiments performed with a wave generator, which produces sinusoidal vertical modes propagating along a rectangular tank. We measured
the growth rate of the instability. This quantity is of paramount importance to single out  
the major mechanism in dissipation processes, a recently highly debated issue~\cite{Kunze,debate,debatebis}.  We first briefly outline theoretical aspects of this instability, after which the experimental configuration is described. Then we present our experimental results and compare some of them with theoretical predictions.

\vspace*{0.5cm}
\paragraph*{Theory}
Internal waves are characterized by the buoyancy frequency, $N=\sqrt{(-g/\bar{\rho_0}) ({\rm d}\rho_0/{\rm d}z)}$, in which $g$ is the acceleration of gravity,  $\bar{\rho_0}$ the characteristic fluid density and $({\rm d}\rho_0/{\rm d}z)$ the density gradient in the vertical direction~$z$.  At large Prandtl number, the 2-D Boussinesq  equations of motion can be written as
\begin{equation}
\frac{\partial^2\nabla^2\psi}{\partial t^2}+N^2\frac{\partial^2\psi}{\partial x^2}=\frac{\partial }{\partial t}J(\psi,\nabla^2\psi) -\frac{g}{\rho_0} \frac{\partial}{\partial x}J(\tilde{\rho},\psi)+\nu \nabla^4\psi_t\,,\label{Boussinesq}
\end{equation}
where $\tilde{\rho}=\rho-\rho_0$ is the perturbation density field, $\psi$ the stream function, $J$ 
  the Jacobian operator  and $\nu$ the viscosity.  Seeking wave solutions with wave number $\overrightarrow{k}=(k,m)$, Eq.~(\ref{Boussinesq}) 
 leads to the inviscid linear dispersion relation for frequency~$\omega$,
\begin{equation} 
\omega^2=N^2\frac{k^2}{k^2+m^2}.\label{dispersion_rel}
\end{equation}
For small amplitudes, it can be assumed that that several waves concurrently exist simply as a linear superposition. In the case of a resonant triad interaction, where 
three waves satisfy the spatial resonance condition
\begin{equation}
\overrightarrow{k_0}=\overrightarrow{k_1}+\overrightarrow{k_2}\,, \label{PSIspatial}
\end{equation}
and the temporal resonance condition
\begin{equation}
{\omega}_0={\omega}_1+{\omega}_2\,,\label{PSItemporal}
\end{equation}
the nonlinear terms of Eq.~(\ref{Boussinesq}) act as forcing terms transferring energy between the three waves. Each  wave
must satisfy the dispersion relation~(\ref{dispersion_rel}). A finite amplitude, large length scale, high frequency wave 
($\overrightarrow{k_0}, \omega_0$)
can transfer energy to produce two secondary waves of smaller length scales and lower frequencies, ($\overrightarrow{k_1}, \omega_1$) and ($\overrightarrow{k_2}, \omega_2$). The instability results from a competition between nonlinear effects and viscous dissipation.  The growth is exponential if the amplitude of the secondary waves is initially small compared to the amplitude of the primary wave~\cite{Koudella2006, McEwan1977}. In this case, the growth rate is equal to
\begin{equation}
\lambda=-\frac{1}{2}(T_1+T_2)+\left[\frac{1}{4}(T_1-T_2)^2+I_1I_2\psi_0^2\right]^{1/2}\,,
\label{growthrate}
\end{equation}
where $\psi_0$ is the amplitude of the stream function of the primary wave, $I_1$ and $I_2$ are the interaction coefficients 
\begin{equation}
I_i=\frac{k_{p}m_{q}-k_{q}m_{p}}{2\omega_i \kappa_i^2}\left[\omega_i(\kappa_p^2-\kappa_q^2)+N^2k_{i}\left(\frac{k_{p}}{\omega_p}-\frac{k_{q}}{\omega_q}\right)\right]
\end{equation}
and $i,p,q = 0,1$ or $2$ while  $T_i=\nu\kappa_i^2/2$ is the viscous damping factor of the wave $i$  and $\kappa^2=k^2+m^2$.

\vspace*{0.5cm}
\paragraph*{Experimental Configuration}
A tank, $160$~cm long and $17$~cm wide, is filled with linearly stratified salt water with constant buoyancy frequency~$N$ using the standard double bucket method.  A  monochromatic vertical mode-1 wave is generated using the wave generator employed in previous experiments~\cite{Gostiaux2007,Mercier2010}. The generator is composed of $50$ plates moving horizontally to impose the horizontal velocity component of a mode-1, {\it i.e}, $u(x=0,z,t)=-a\omega_0\cos(\pi z/H)\cos(\omega_0 t)$, $H$~being the water depth, $\omega_0$ the excitation frequency and $a$ the amplitude of the oscillation of the plates. The motion of the fluid is captured by the synthetic schlieren technique using a dotted image behind the tank~\cite{Dalziel00}. A camera is used to acquire images of this background at    $1.875$ frames per second. The CIVx algorithm~\cite{Fincham2000}  computes the cross-correlation between the real-time and the $t=0$ background images,  giving the variation of the horizontal, $\tilde{\rho}_x(x,z,t)=\partial_x(\rho(x,z,t)-\rho_0(z))$, and vertical, $\tilde{\rho}_z(x,z,t)=\partial_z(\rho(x,z,t)-\rho_0(z))$,  density gradients.
\begin{figure}[htb]
\includegraphics[width=\linewidth]{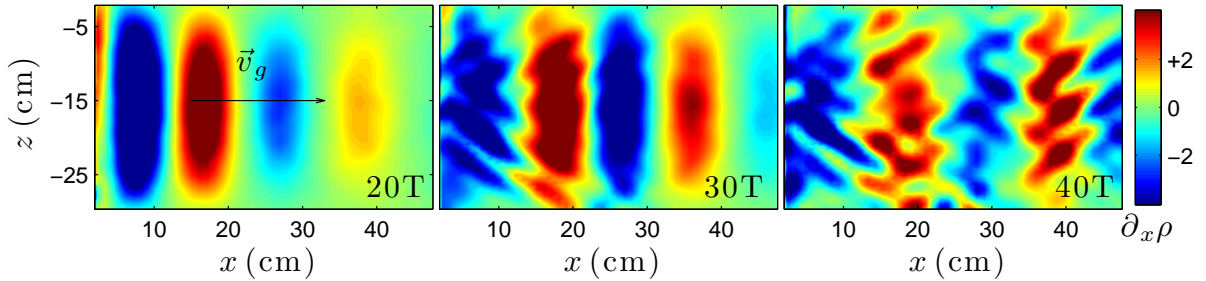}
\caption{(Color online)  Snapshot of the horizontal density gradient (in kg/m$^{4}$) obtained at $t=20T$, $t=30T$ and $t=40T$ 
with the parameters $a=0.5$~cm,  $\omega_{0}=0.95N$ and $N=0.822$~rad$\cdot$ s$^{-1}$. On the left-hand panel the direction of the group velocity, $\vec{v}_g$, is indicated. After $20T$, the primary wave has only reached a steady state in the first 20 cm from the generator (see Fig.~\ref{spectra}).}
\label{PSI}
 \end{figure}

\begin{figure}[htb]
\includegraphics[width=\linewidth]{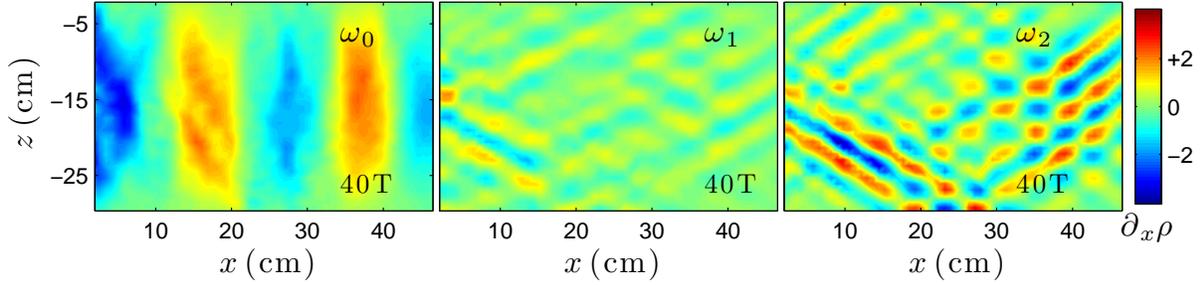}
\caption{(Color online) Real part of the Hilbert Transform  of $\tilde{\rho}_x$  at $t=40T$ (see Fig.~\ref{PSI}(right)) presented after filtering at $\omega_{0}=0.95N$ (left), $\omega_{1}=0.38N$ (center) and $\omega_{2}=0.57N$ (right). Note that the color scale is the same as in Fig.~\ref{PSI}.} 
\label{PSIbis}
\end{figure}

\begin{figure}[htb]
\includegraphics[width=0.6\linewidth]{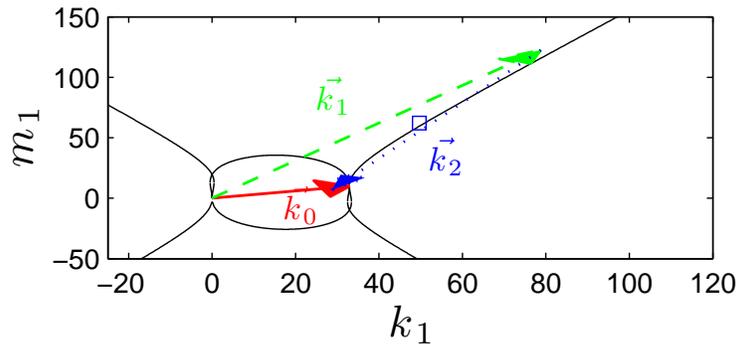}
\caption{(Color online) Spatial resonance conditions for the experiment presented in Fig.~\ref{PSI}. The black line represents the theoretical location of the tip of the wave vector $\vec{k_1}$ for a given primary wave vector~$\vec{k_0}$ so that the resonance conditions~(\ref{PSIspatial}) and~(\ref{PSItemporal}) and dispersion relations~(\ref{dispersion_rel}) are satisfied. The three arrows are the experimental measurement of the three wave vectors: the red solid arrow is the primary wave vector~$\vec{k_0}$, the green dashed and blue dotted arrows the secondary wave vectors~$\vec{k_1}$ and $\vec{k_2}$. The square $\square$ represents the most unstable theoretical mode.} \label{triad}
\end{figure}

\vspace*{0.5cm}
\paragraph*{Results}
Snapshots of an experimental horizontal density gradient field at different times for  a particular experiment are presented in Fig.~\ref{PSI}. At early times, a pure vertical mode-1  wave can be seen propagating to the right away from the wave generator located at $x=0$: this is the {\it  primary wave.} After several  frequency periods $T$ (typically 30), this wave is destabilized and two {\it  secondary waves} appear, with different frequencies and wave numbers from the primary wave. To see these waves more clearly,  the horizontal density gradient at later times is filtered at the frequency of the primary wave, $\omega_0$ and at the frequencies of the two secondary waves $\omega_1$ and $\omega_2$.  As described below, the frequencies $\omega_1$ and $\omega_2$ were determined from a power spectrum. The result is shown in Fig.~\ref{PSIbis}. 
Some of the energy of the primary wave has been transferred to both secondary waves, leading to a decrease in the amplitude of the primary wave (compare the left part of Fig.~\ref{PSI}(left) and Fig.~\ref{PSIbis}(left)). These two waves have smaller frequency and also smaller wavelength. In agreement with the dispersion relation, which links the frequency to the angle of propagation of the wave, the angle of constant phase is different for the two wavelengths. For the experiment presented in Fig.~\ref{PSI}, the three measured frequencies $(\omega_0,\omega_1,\omega_2)$ are equal to $(0.95,0.38,0.57)N$, attesting that the temporal resonance condition~(\ref{PSItemporal}) is satisfied. To justify that the spatial resonance condition~(\ref{PSIspatial}) is also satisfied, the components of the three wavevectors have to be measured. This is done by extracting the phase of the signal at a given frequency, $\omega_{0,1,2} t\pm k_{0,1,2}x\pm m_{0,1,2}z$, using a Hilbert transform~\cite{HilbertTransform}. At a fixed time and $x$ (respectively $z$), the phase is linear with the position $z$ (resp. $x$). The component $m_{0,1,2}$ (resp. $k_{0,1,2}$) is given by the slope of the linear fit. Within experimental errors, the wave vectors, represented in Fig.~\ref{triad}, satisfy the theoretical spatial resonance condition~(\ref{PSIspatial}).

The measured density  gradient fields are then analyzed using a time-frequency representation calculated at each spatial point
\begin{equation}
S_x(t,\omega)=\left|\int_{-\infty}^{+\infty}{\rm d} u\, \tilde{\rho}_x(u)\, e^{ i \omega u}\, h(t-u)\right|^2\,,\label{Timefreqeq}
\end{equation}
where $h$ is a smoothing Hamming window of energy unity~\cite{Flandrin99}.   Good frequency resolution is provided by a large time window $h$ while good time resolution is provided by small time window $h$.
To increase the signal to noise ratio, the data is averaged along a vertical line over the water depth. For large $\omega_0/N$ values, the dissipation length is small, so the analysis line is chosen to be close to the generator so that the amplitude is large.

\begin{figure}[htb]
\includegraphics[width=\linewidth]{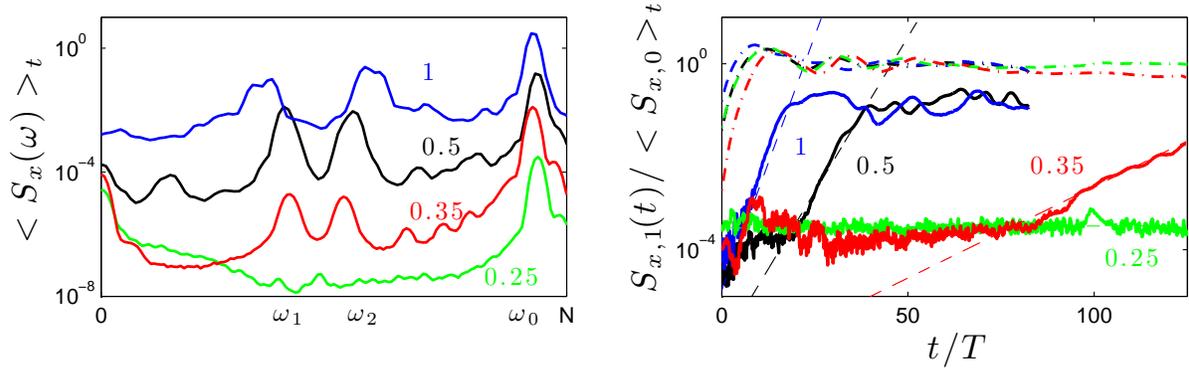}
\caption{(Color online) (Left) Spectra of the density  gradient field, $\langle S_{x}(\omega)\rangle_t$, for four different excitation amplitudes for $\omega_0=0.94N$ measured at $x\approx12$~cm. The amplitudes are respectively  $0.25$~cm (green), $0.35$ (red), $0.5$ (black) and $1$ (blue). The value of the amplitude is indicated next to the corresponding curve. The spectra 0.25, 0.35 and 1 are multiplied respectively by $0.01$, $0.1$ and $10$ for  illustration purposes. (Right) Amplitude of the secondary wave~$\omega_1$, $S_{x,1}(t)$, normalized by the amplitude of the primary wave, $\langle S_{x,0}\rangle_t$, averaged over time when the steady-state of the mode-1 has been reached. Results are similar for the other secondary wave~$\omega_2$. The dashed-dotted lines represent the amplitude of the primary wave using the same normalization.
The dashed lines  shows the linear fit, which gives the value of the growth rate $\lambda$. } \label{spectra}
 \end{figure}

Fig.~\ref{spectra}(left) presents the spectra of the density field for four different excitation amplitudes with $\omega_0=0.94N$.  The spectra are obtained using a time window width equal to $100$~T to have good frequency resolution. $S_x(t, \omega)$ is then averaged over the $10$ last periods. Analyzing first the result for the amplitude $0.5$  cm, the picture emphasizes a large peak close to $N$, corresponding to the frequency of the mode-1  wave. 
A pair of twin peaks are observed, corresponding to secondary waves of frequencies, $\omega_1$ and $\omega_2$, smaller than $\omega_0$.

The  amplitude of each wave  is then computed using a time-frequency analysis with a time window width equal to $20$~T to increase time resolution. The amplitude of the secondary wave of frequency $\omega_1$ is presented in Fig.~\ref{spectra}(right). After several forcing periods, a steady state for the primary wave is reached. After a time interval, the secondary wave starts to grow and a linear increase of the amplitude on a semilogarithmic plot is observed, confirming    exponential growth. The value of the growth rate~$\lambda$ is measured using a linear fit, shown with the dashed lines in Fig.~\ref{spectra}(right). The amplitude of the secondary waves eventually saturates. 
 
Comparing the different curves in Fig.~\ref{spectra}, one observes that the amplitude has an influence not only on the location but also on the height of the peaks of the secondary waves in the spectrum.  If the amplitude of the primary wave is too small, no peaks are visible and therefore no instability  is observed  during the experiment run time, $T_{\rm run}$. This result shows that the growth rate in this particular case has to be smaller than 
$1/{\rm run}$. It may also give an indication of the existence of a threshold in amplitude.   As the amplitude increases, the distance between the two peaks increases and the instability occurs earlier (after fewer forcing periods) and   is stronger, {\it i.e.} with a larger growth rate which is in agreement with the theoretical growth rate~(\ref{growthrate}).
\begin{figure}[htb]
\includegraphics[width=\linewidth]{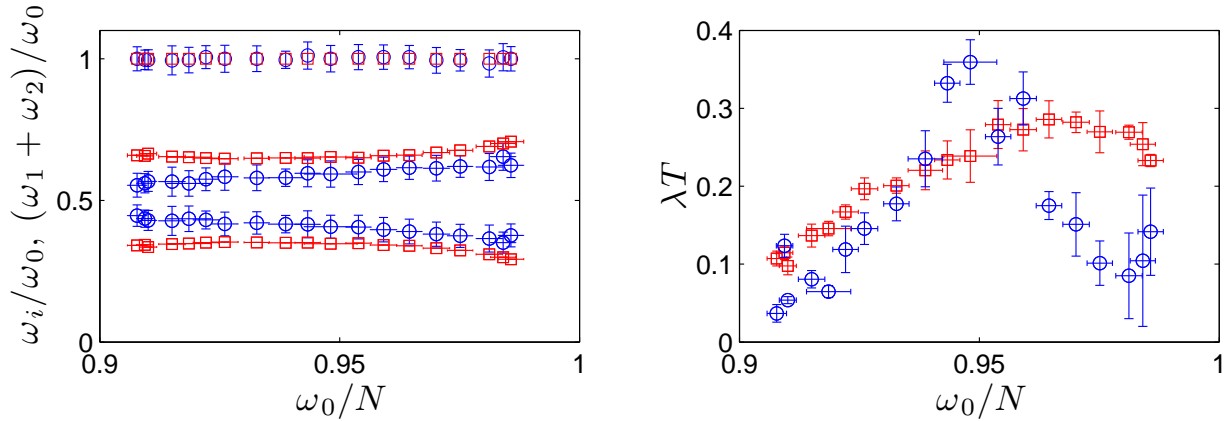}
\caption{(Color online) (Left) Values of the frequencies of the secondary waves for the experiments (blue $\circ$) and the theoretical calculation (red $\square$). The error bars are the half-width of the corresponding peak (vertical error bars are smaller than the symbols). The top line of point corresponds to the sum of the frequencies, with corresponding colors for theoretical and experimental data. (Right) Values of the growth rate of the secondary waves for the experiments (blue $\circ$) and the theoretical calculation (red $\square$).  Vertical error bars for the experimental results come from the choice of the horizontal location and from the linear fit. These values are smaller than the symbol. $T$ is the period of the primary wave. The theoretical values ($\square$) are computed using Eq.~(\ref{growthrate}) and the measured amplitude of the primary wave. The vertical error bars come from the error made measuring the amplitude of the mode-1.}\label{omega}
\end{figure}

Experiments were performed using the same stratification and an amplitude of $0.5$~cm for frequencies in the range of $0.9<\omega_0/N<1$. For each experiment, the value of the frequencies of the two secondary waves, $\omega_1$ and $\omega_2$, and the growth rate $\lambda$ were measured.  Experimental results are presented as a function of the frequency of the primary wave, $\omega_0/N$, in Fig.~\ref{omega}. The sum of the frequencies of the two secondary waves, $\omega_1 + \omega_2$, is equal to the frequency of the primary wave, $\omega_0$, within experimental errors, in agreement with Eq.~(\ref{PSItemporal}). As $\omega_0/N$ increases, the distance between the two secondary frequencies is larger. The measured value of the growth rate is presented in Fig.~\ref{omega}(right). The growth rate increases to reach a maximum around  $\omega_0 = 0.95N$ and then decreases as $\omega_0$ gets closer to $N$.

To compare quantitatively the experimental results with the theoretical prediction of the growth rate, 
the value of the amplitude of the mode-1  wave  has to be precisely known. The theoretical value of 
the amplitude of the streamfunction is equal to $a \omega_0/m_0$. However, the conversion efficiency from the energy of the wavemaker to the energy of the mode-1 is less than unity and 
depends on experimental conditions~\cite{Mercier2010}. Moreover, as $\omega_0$ gets closer to the cut-off frequency, $N$, the value of the viscous 
damping increases~\cite{Echeverri09}. Consequently, the efficiency is not the same for all primary frequencies~$\omega_0$,
and the amplitude of the primary wave has to be measured experimentally to compute the theoretical value of the growth rate. 
 It is important to check that the steady-state of the mode-1 wave has been reached. However, the tank being finite in length, the measurement has to be performed before the mode-1 wave reflects back into the measurement area.
Then, using a linear polarization relation, the amplitude $\psi_0$ of the stream function at this particular frequency and wave number is
$\psi_0 = {g\omega_0}\partial_x \tilde\rho_0/(4{k_0^2\bar{\rho}N^2})$.
The  theoretical frequency pair ($\omega_1$,$\omega_2$) of the instability is defined as the one that maximizes the growth rate. Without adjustable parameters, the comparison between  experimental and  theoretical results, presented in Fig.~\ref{omega}, emphasizes a good quantitative agreement.

\vspace*{0.5cm}
\paragraph*{Conclusions}
We have reported the first experimental measurement of the growth rate of parametric subharmonic instability in stratified fluids and   we have demonstrated this effect in a systematic set of laboratory experiments   allowing careful comparisons with
theoretical predictions. In practice, this    heavily debated  mechanism~\cite{debate} has implications for many geophysical scenarios.
Interestingly, although the generation mechanisms of oceanic IGW are quite well understood, the comprehension of the processes by which they dissipate is much more open. Consequently, determining the relative importance of parametric subharmonic instability, among the four
recognized dissipation processes~\cite{Kunze}, is the next step in furthering our understanding of
how internal waves impact ocean mixing. Quantitative measurements of the subsequent mixing together with  a fundamental
study of    wave turbulence would be of high interest.   

\acknowledgments
The authors thank G. Bordes, P. Borgnat, B. Bourget, C. Staquet,  for helpful discussions.
This work has been partially supported by the PIWO grant (ANR-08-BLAN-0113-01) and the ONLITUR grant (ANR-2011-BS04-006-01). This work has been partially achieved thanks to the ressources of PSMN (P\^ole Scientifique de Mod\'elisation Num\'erique) de l'ENS de Lyon.

\end{document}